\documentclass[a4paper,fleqn,usenatbib]{mnras}

\usepackage{newtxtext,newtxmath}

\usepackage[T1]{fontenc}
\usepackage{ae,aecompl}

\usepackage{graphicx}	

\newcommand{\kms}{km\,s$^{-1}$}

\newcommand{\figps}[3]{\resizebox{#1}{!}{\rotatebox{#2}{\includegraphics{#3}}}}

\title[Eclipsing HgMn stars]{New eclipsing binaries with mercury-manganese stars}

\author[O. Kochukhov et al.]
{O.\ Kochukhov$^1$\thanks{E-mail: oleg.kochukhov@physics.uu.se},
J.\ Labadie-Bartz$^2$,
V.\ Khalack$^3$,
M.\ E.\ Shultz$^4$,
\\
$^1$Department of Physics and Astronomy, Uppsala University, Box 516, Uppsala 75120, Sweden \\
$^2$Instituto de Astronomia, Geof\'isica e Ciencias Atmosf\'ericas, Universidade de S\`ao Paulo, Rua do Mat\~ao 1226,
Cidade Universit\~aria, \\S\'ao Paulo, SP 05508-900, Brazil\\
$^3$D\'epartement de Physique et d'Astronomie, Universit\'e de Moncton, Moncton, NB E1A 3E9, Canada\\
$^4$Department of Physics and Astronomy, University of Delaware, 217 Sharp Lab, Newark, Delaware, 19716, USA
}

\date{Accepted 2021 June 11. Received 2021 June 8; in original form 2021 May 24}

\pubyear{2020}

\begin{document}
\label{firstpage}
\pagerange{\pageref{firstpage}--\pageref{lastpage}}
\maketitle

\begin{abstract}
Eclipsing binary stars are rare and extremely valuable astrophysical laboratories that make possible precise determination of fundamental stellar parameters. Investigation of early-type chemically peculiar stars in eclipsing binaries provides important information for understanding the origin and evolutionary context of their anomalous surface chemistry. In this study we discuss observations of eclipse variability in six mercury-manganese (HgMn) stars monitored by the TESS satellite. These discoveries double the number of known eclipsing HgMn stars and yield several interesting objects requiring further study. In particular, we confirm eclipses in HD\,72208, thereby establishing this object as the longest-period eclipsing HgMn star. Among five other eclipsing binaries, reported here for the first time, HD\,36892 and HD\,53004 stand out as eccentric systems showing heartbeat variability in addition to eclipses. The latter object has the highest eccentricity among eclipsing HgMn stars and also exhibits tidally induced oscillations. Finally, we find evidence that HD\,55776 may be orbited by a white dwarf companion.
\end{abstract}




\section{Introduction}
\label{sec:intro}

Mercury-manganese (HgMn) stars are late-B main sequence objects distinguished by overabundance of heavy elements, slow rotation, and lack of strong magnetic fields. These stars exhibit some of the most extreme departures from the solar abundance pattern and relative isotope composition of heavy elements \citep*[e.g.][]{ghazaryan:2018}, believed to be produced by radiatively driven atomic diffusion \citep{michaud:2015}. HgMn stars are frequent members of close binaries \citep*{gerbaldi:1985}. These systems are important objects for constraining the origin and long-term evolution of chemical peculiarities. 

Eclipsing binaries (EBs) are particularly valuable sources of model-independent information on stellar radii and masses \citep[e.g.][]{torres:2010} and well-constrained ages \citep{tkachenko:2020}. Despite an abundance of close spectroscopic binaries with HgMn components \citep{ryabchikova:1998a,catanzaro:2004}, only four such objects, HD\,10260, 34364, 267564, and TYC 455-791-1 \citep[][and references therein]{kochukhov:2021a,paunzen:2021a}, were known in eclipsing systems. Eclipses in two more spectroscopic binaries with HgMn primaries, HD\,72208 and HD\,161701, suspected based on low-quality photometric measurements by the STEREO satellite \citep{wraight:2011}, are yet to be confirmed by independent observations. Here we aim to expand the list of EBs with HgMn components and find targets suitable for detailed spectroscopic follow-up  by leveraging the power of high-precision space photometry data.

\section{Observations and data analysis}
\label{sec:obs}

More than 500 confirmed and candidate HgMn stars are currently known (\citealt{renson:2009,chojnowski:2020}; \citealt*{paunzen:2021}; \citealt{gonzalez:2021}). We are carrying out a systematic investigation of variability of these stars using high-precision photometric data acquired by the Transiting Exoplanet Survey Satellite \citep[TESS,][]{ricker:2015}. This mission has surveyed most of the sky during its first two years (2018--2020) of operation. These observations consisted of 26 pointings, also known as sectors, each covering a segment of the Northern or Southern ecliptic hemisphere extending from one of the ecliptic poles down to $\pm$6\degr\ of the ecliptic equator. Each $24\degr\times96\degr$ sector was monitored nearly continuously for 27.4~d. A subset of targets were observed in 2-min cadence mode, with fully reduced and calibrated light curves provided by the TESS science team \citep{jenkins:2016}. These light curves are available for 65 HgMn stars. Detailed analysis of these data will appear in a separate publication (Kochukhov et al., in preparation). Images of entire sectors are available every 30 min. These full-frame images (FFI) have been processed by \citet{huang:2020} to provide uncorrected quick look pipeline (QLP) light curves for about 24.4 million stars. This includes the majority of HgMn stars not observed in 2-min cadence. In this study we used the QLP light curves corresponding to TESS sectors 1--26 to search for EBs among HgMn stars. This search yielded 15 candidates exhibiting eclipse-like light curve behaviour. For all these targets we carried out our own FFI image analysis to confirm that the eclipse signal is associated with an HgMn star. This analysis was extended to data from sectors 33 and 34 for targets confirmed to be EBs and reobserved by TESS during the third year of the mission.

The TESS FFIs were analysed for each of the 15 candidates using the \textsc{lightkurve} package \citep{Lightkurve2018} and \textsc{TESScut} \citep{Brasseur2019} to download a target pixel file (TPF) with a 50 $\times$ 50 pixel grid centered on the target star's coordinates for every available TESS sector. An aperture was chosen to limit contamination from neighboring sources (generally being smaller than what was used in the QLP), and standard PCA detrending was applied. Light curves were also generated for each pixel in the FFIs, allowing us to determine which specific pixels carry the EB signals. This analysis confirmed eclipses in six of the candidate systems, while in the other nine the eclipses originate in a neighboring source.

\begin{table*}
\caption{Characteristics of the eclipsing binary stars discussed in this study. The columns give the HD or TYC number, the TIC number, $V$ magnitude, TESS sectors in which targets were observed, the number of primary and secondary eclipses in the analysed light curve, the derived orbital period and eccentricity, and the spectroscopic binary classification when available. \label{tab:results}}
\begin{tabular}{llllllll}
\hline
Name & TIC & $V$ (mag) & TESS sectors & $N_{\rm p},N_{\rm s}$ & $P_{\rm orb}$ (d) & $e$ & SB \\
\hline
TYC 4047-570-1 & 50517575 & 11.19 & 18 & 6, 8 & 3.275965(10) & 0.0 & $\cdots$ \\
HD 36892 & 115637849 & 8.79 & 19 & 3, 3 & 8.4590(27) & 0.25 & SB1 \\
HD 50984 & 124091234 & 8.14 & 6, 7, 33 & 9, 9 & 7.1637811(24) & 0.0 & SB1 \\
HD 55776 & 178480331 & 10.01 & 7, 33 & 3, 0 & 18.09724(38) & $\cdots$ & SB1 \\
HD 53004 & 291583988 & 7.32 & 7, 33 & 4, 4 & 11.538878(50) & 0.71 & $\cdots$ \\
HD 72208 & 366577510 & 6.80 & 7, 34 & 2, 2 & 22.01162(28) & 0.31 & SB2 \\
\hline
\end{tabular}
\end{table*}

We measured the times and durations of primary and secondary eclipses for all confirmed eclipsing HgMn stars. These measurements were used to obtain binary periods by fitting straight lines to centroids of eclipses. In addition, we estimated orbital eccentricities with the method described by \citet{prsa:2018}.

\section{Results}
\label{sec:res}

\subsection{New and confirmed eclipsing binaries}

\begin{figure*}
\centering
\figps{0.95\hsize}{0}{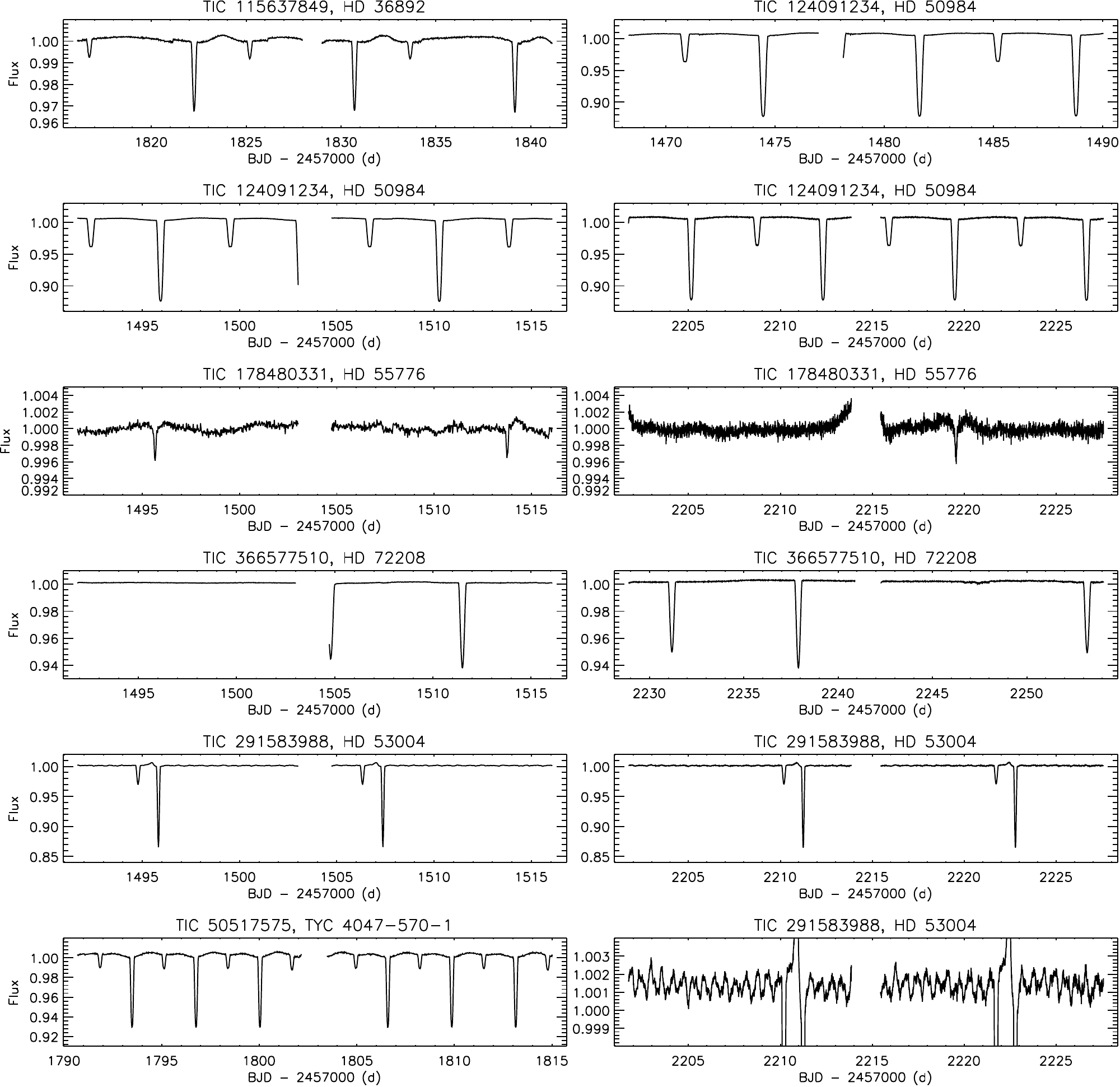}
\caption{TESS light curves of HgMn stars. The bottom-right panel zooms on oscillations in the sector 33 light curve of HD\,53004.}
\label{fig:lcs}
\end{figure*}

Based on the pixel-by-pixel time series analysis described above we have established coincidence of eclipses with six HgMn stars. Their properties and the results of the light curve analyses are summarised in Table~\ref{tab:results}. The light curves themselves are presented in Fig.~\ref{fig:lcs}.

\textbf{TYC 4047-570-1 (TIC 50517575)} was observed in TESS sector 18. The light curve shows eclipses as well as an ellipsoidal variation. We determined $P_{\rm orb}=3.275965$~d based on the measurement of times of the six primary and eight secondary eclipses. The relative timing of the primary and secondary eclipses is consistent with a circular orbit. This $V=11.2$ mag star was classified as an HgMn object by \citet{chojnowski:2020} using APOGEE spectroscopy in the H band. A single observation was obtained by these authors, which does not allow one to investigate radial velocity variation. At the same time, the projected rotational velocity reported for this star, $v_{\rm e}\sin i = 43\pm9$~\kms, is consistent with synchronous rotation of a late-B star in a short-period binary. Under this assumption and using $i=90\degr$ one obtains $R=2.78\pm0.58$~$R_\odot$.

\begin{figure*}
\centering
\figps{0.32\hsize}{0}{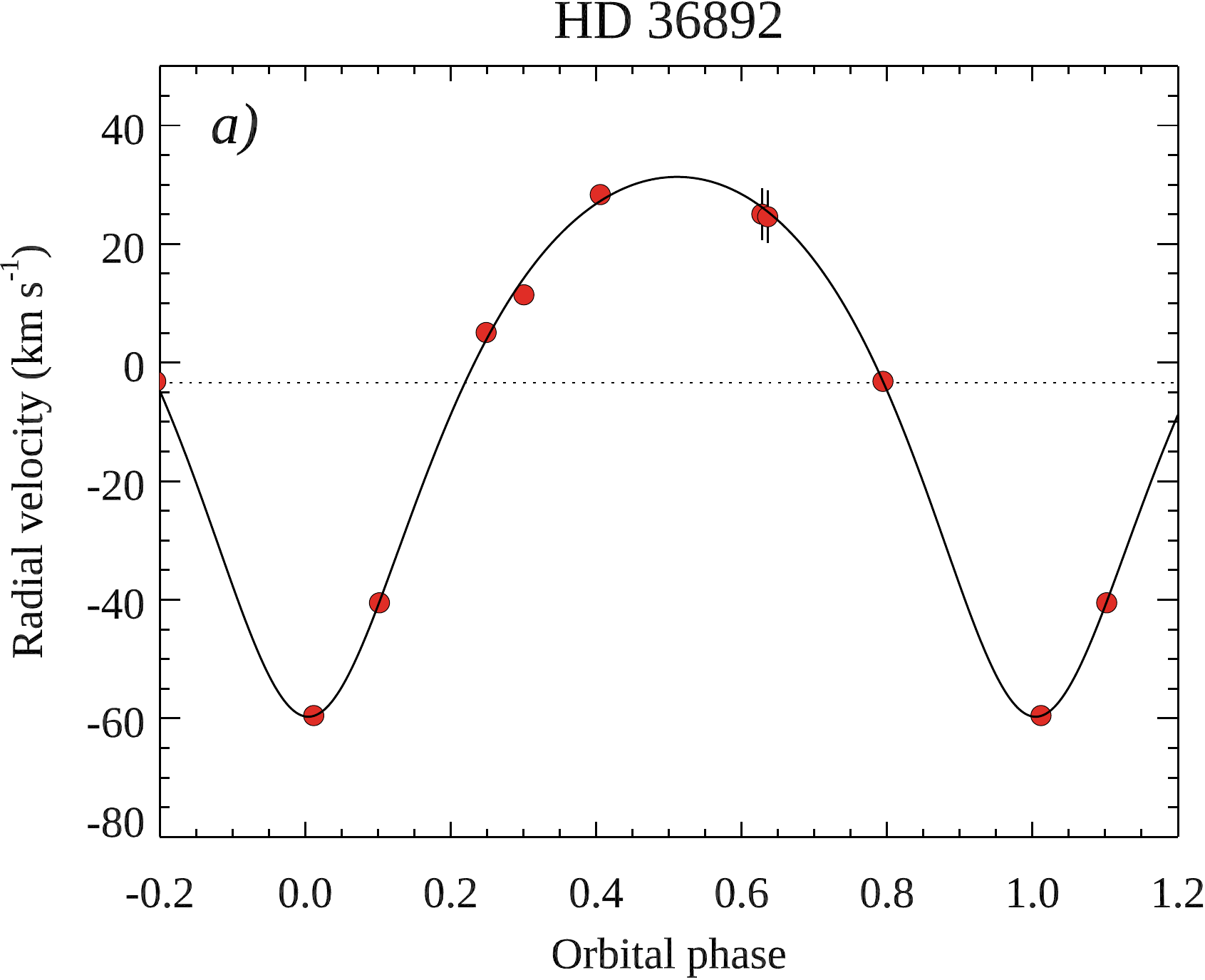}\hspace*{3mm}
\figps{0.32\hsize}{0}{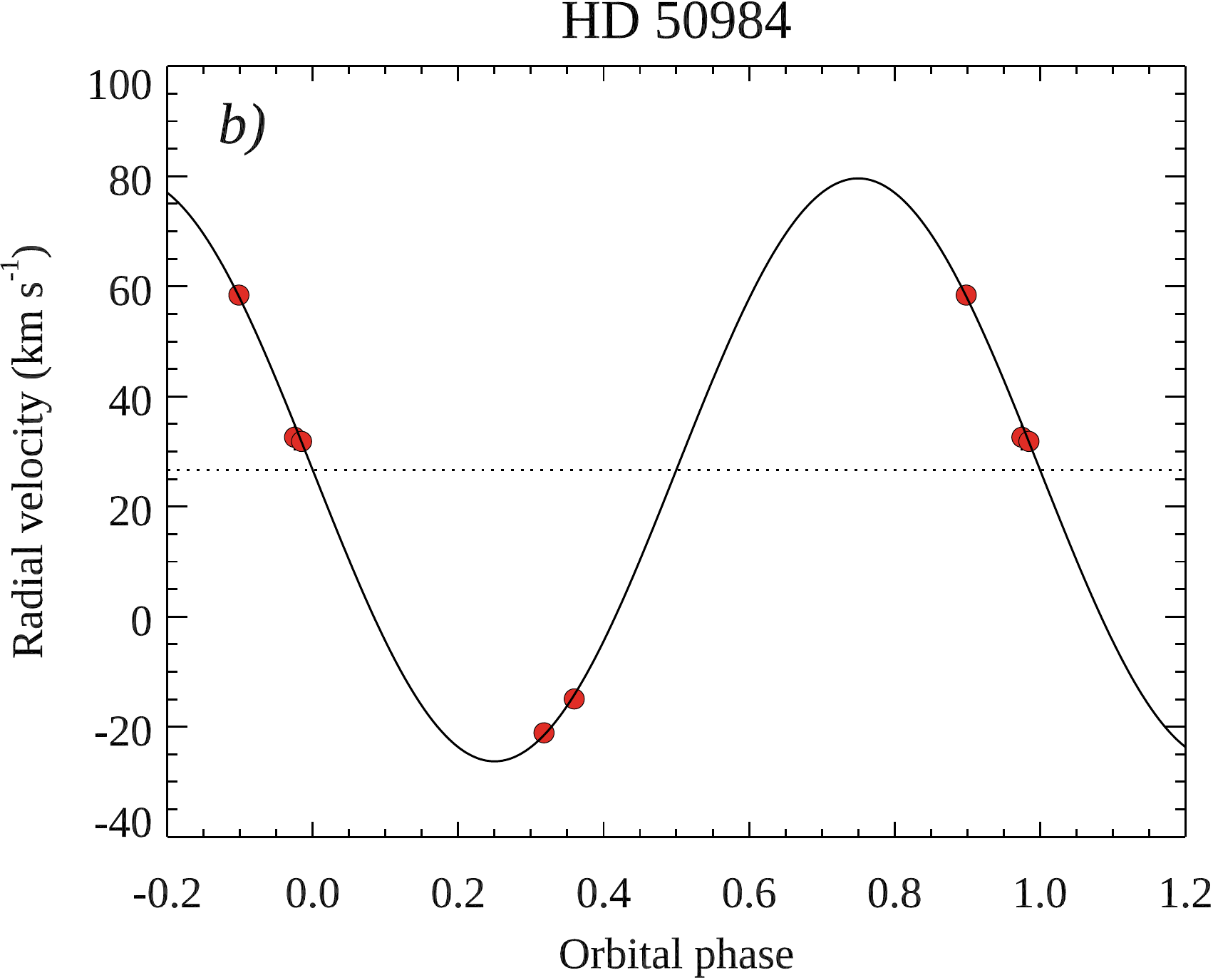}\hspace*{3mm}
\figps{0.32\hsize}{0}{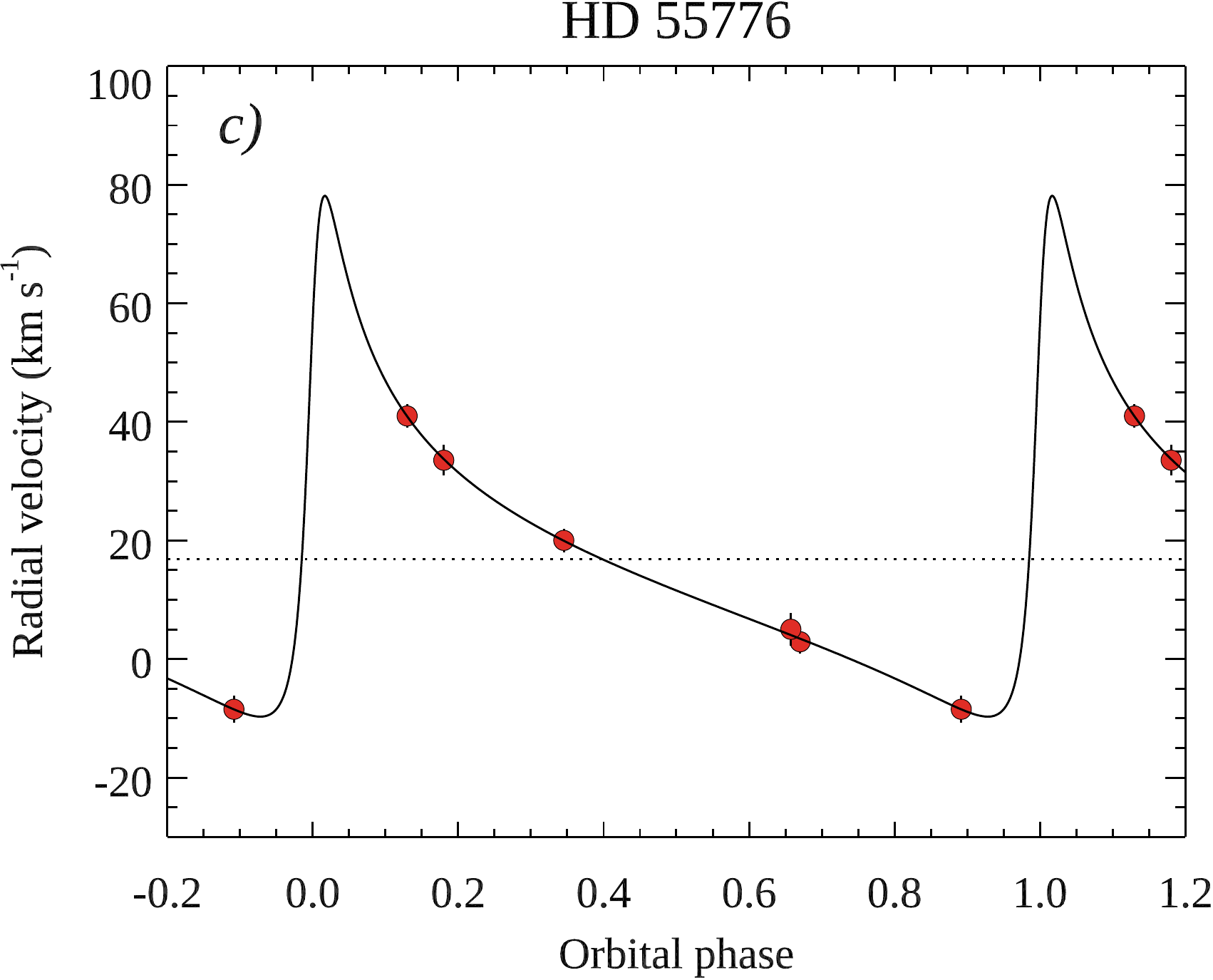}
\caption{Spectroscopic orbital solutions for the SB1 systems HD\,36892 (a), HD\,50984 (b), and HD\,55776 (c). Symbols show radial velocity measurements by \citet{chojnowski:2020}. The solid line corresponds to orbital fits discussed in the text.}
\label{fig:rv1}
\end{figure*}

\textbf{HD 36892 (TIC 115637849)} exhibits eclipses accompanied by an ellipsoidal and heartbeat variation. This target was observed by TESS in sector 19 with three primary and three secondary eclipses visible in the light curve. We determined $P_{\rm orb}=8.4590$~d and $e=0.25$ from the TESS data. \citet{chojnowski:2020} have classified this $V=8.8$ mag object as an SB1 HgMn star and derived a spectroscopic orbital solution with a period of 16.109~d and $e=0.32$ using eight radial velocity measurements. This orbital solution appears to be spurious since the shorter TESS period phases available radial velocity measurements equally well. Fig.~\ref{fig:rv1}a illustrates our revised orbital fit to the radial velocities by \citet{chojnowski:2020}. A good description of the spectroscopic measurements is obtained with the following orbital elements: $P_{\rm orb}=8.4567\pm0.0013$~d, $HJD_0=2456667.32\pm0.24$, $K_1=45.5\pm2.1$~\kms, $\gamma=-3.4\pm2.9$~\kms, $e=0.24\pm0.06$, and $\omega=87\pm13\degr$. Both $P_{\rm orb}$ and $e$ of this revised spectroscopic orbital solution are consistent with the orbital parameters inferred from the TESS light curve.

\textbf{HD 50984 (TIC 124091234)} has the longest TESS data set. This star was observed in sectors 6, 7 and then again in sector 33. Its light curve shows nine primary and the same number of secondary eclipses. The orbit is circular and there is a weak ellipsoidal variation. Analysis of eclipse timings yields $P_{\rm orb}=7.1637811$~d. A relatively bright star ($V=8.1$ mag), HD\,50984 was attributed the HgMn classification by \citet{chojnowski:2020}. Based on five observations, these authors reported this star to be an SB1 system with a large radial velocity variation but did not provide a spectroscopic orbital solution. Assuming a circular orbit and fixing the orbital period to the precise photometric value, we found $HJD_0=245598.618\pm0.029$, $K_1=52.9\pm1.6$~\kms, and $\gamma=26.66\pm0.86$~\kms. This orbital solution is shown in Fig.~\ref{fig:rv1}b. The projected rotational velocity of the primary, $v_{\rm e}\sin i=15\pm4$~\kms\ \citep{chojnowski:2020}, corresponds to a stellar radius of $2.12\pm0.57$~$R_\odot$ assuming a synchronous stellar rotation and equator-on geometry.

\textbf{HD 55776 (TIC 178480331)} exhibits an unusual eclipse variability in TESS data. Three eclipses, corresponding to a period of 18.09724~d are evident in the light curves observed in sectors 7 and 33. Only one set of eclipses is seen. These eclipses are faint and short in duration, suggesting a faint and small companion. At the same time, the light curve shows an increase in brightness around the eclipses. If this is interpreted as an illumination effect on the primary, the secondary must be a hotter object. These characteristics are compatible with a white dwarf companion. 

\citet{chojnowski:2020} reported HD~55776 to be an SB1 system with significant radial velocity variation and a period of 22.69~d. This orbital solution, based on only six radial velocity measurements, is not unique. Using the new photometric period, we are able to fit the same radial velocity data with 
a broad range of eccentric orbits having $K_1=49\substack{+30 \\ -18}$~\kms, $e=0.82\substack{+0.09 \\ -0.15}$, and $\omega=200\pm20\degr$ (see Fig.~\ref{fig:rv1}c for a representative orbit with parameters within this range). A more accurate orbital solution cannot be derived due to a poor coverage of periastron.
Further observations are required to improve the orbital phase coverage. 

\begin{figure}
\centering
\figps{0.9\hsize}{0}{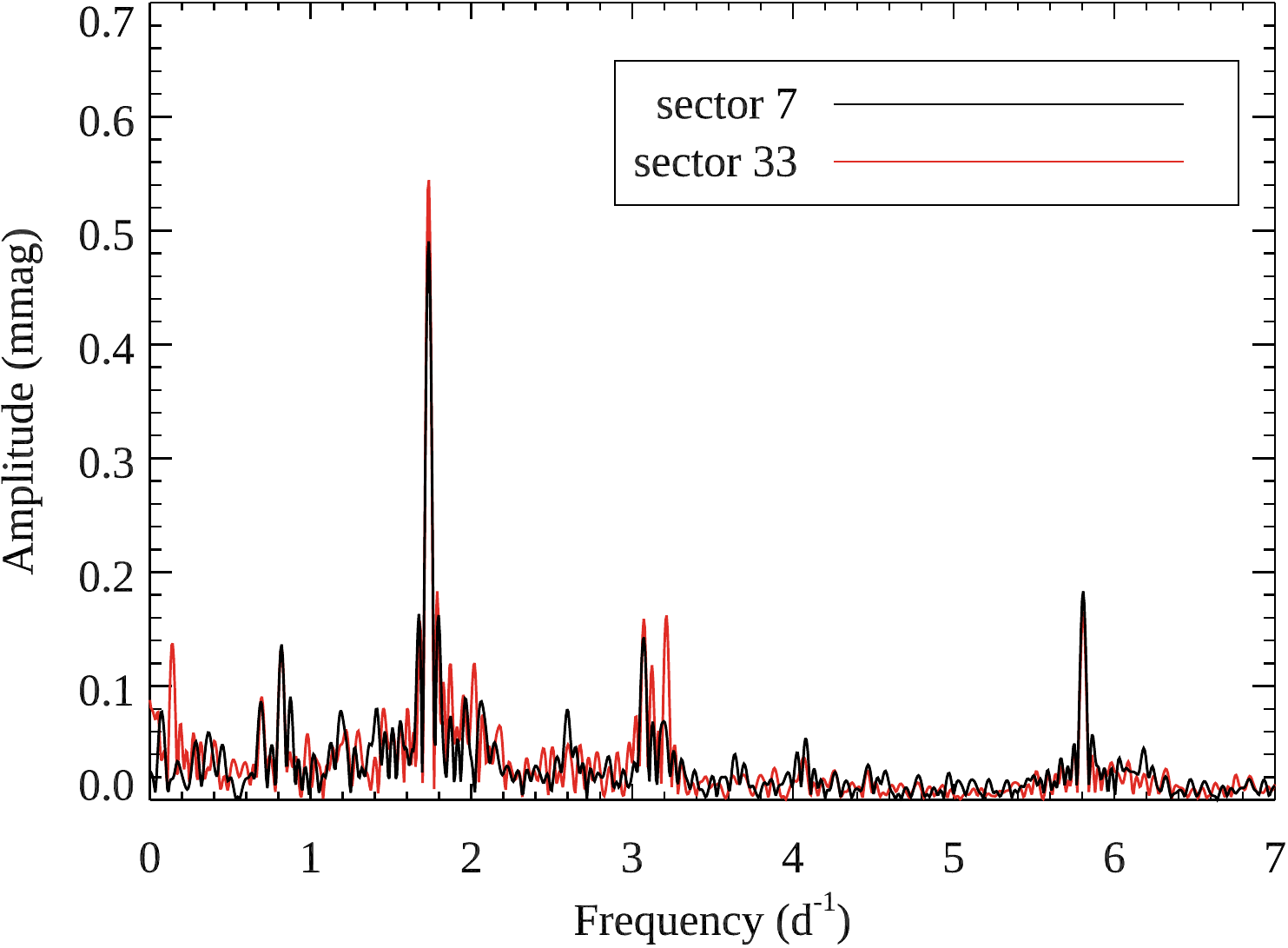}
\caption{Amplitude spectrum of the tidally induced oscillations in the TESS light curve of HD\,53004.}
\label{fig:puls}
\end{figure}

\textbf{HD 53004 (TIC 291583988)} is the second-brightest object ($V=7.3$ mag) in our study. Its TESS light curve from sectors 7 and 33 reveals a highly eccentric eclipsing system with a heartbeat variation. Using four pairs of primary and secondary eclipses, we found $P_{\rm orb}=11.538878$~d and $e=0.71$. In addition, multi-periodic, likely tidally induced, oscillations are visible in the TESS light curve. According to our frequency analysis, at least three pulsation signals, with periods of 0.172, 0.326, and 0.577~d, are present in both sectors (Fig.~\ref{fig:puls}). The latter period is an integer fraction of the orbital period, confirming that it corresponds to a tidally induced pulsation.

HD\,53004 was classified as an HgMn star by \citet{niemczura:2009}. They have determined $T_{\rm eff}=11600$~K, $\log g=4.0$ together with abundances of several chemical element and deduced a moderately high projected rotational velocity of $58\pm8$~\kms. An alternative analysis by \citet{lefever:2010} yielded $T_{\rm eff}=11000$~K, $\log g=3.9$, and $v_{\rm e}\sin i=51\pm7$~\kms. Both studies used the same FEROS spectrum of HD\,53004 and did not assess radial velocity variability. This star is known to have a 0.5~mag fainter visual companion separated from the primary by 0\farcs1 \citep{mason:2001}. This companion has an orbital period of many decades and is therefore not part of the 11.5~d eclipsing binary discovered here. However, it can still contribute to the TESS light curve and high-resolution spectrum of HD\,53004.

\textbf{HD 72208 (TIC 366577510, HR 3361)} is the brightest object ($V=6.8$ mag) in this study. It was observed by TESS in sector 3, when only the primary eclipse was well-covered. The second visit in sector 34 provided coverage of one primary and two secondary eclipses. Considering the timing of both primary and secondary eclipses, we found $P_{\rm orb}=22.01162$~d and $e=0.31$. This star was previously reported as a candidate eclipsing binary by \citet{wraight:2011}, who derived $P_{\rm orb}=22.0130$~d and $|e\times\cos{\omega}|=0.391$, which is close to our results. 

\begin{figure}
\centering
\figps{0.9\hsize}{0}{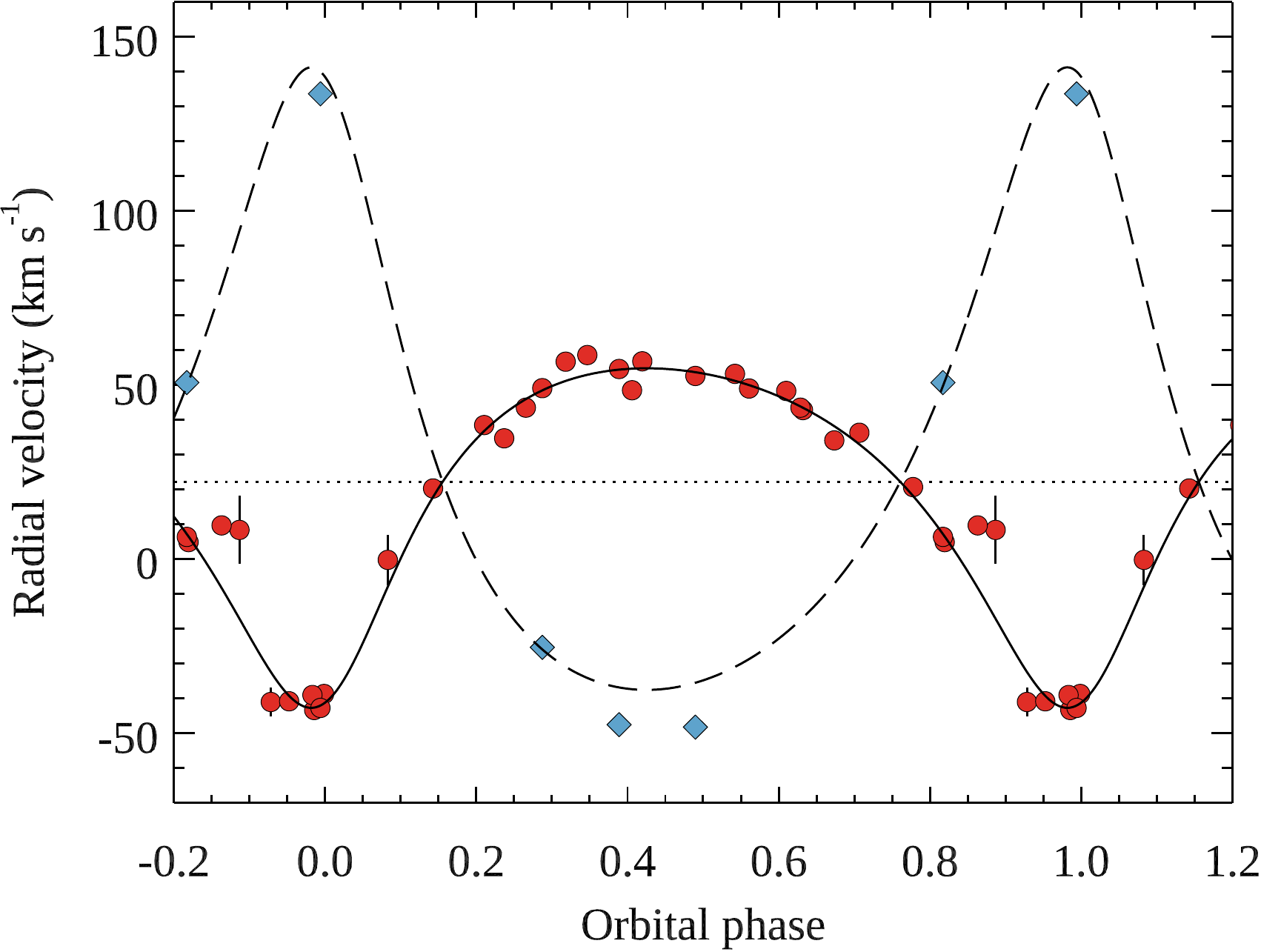}
\caption{Spectroscopic orbital solution for the SB2 system HD\,72208. Symbols show radial velocity measurements by \citet{stickland:1984} for the primary (circles) and secondary (diamonds). The solid and dashed lines correspond to the orbital fit for the primary and secondary respectively.}
\label{fig:rv4}
\end{figure}

HD\,72208 is the only star in our survey previously known to be an SB2 system. Using radial velocity measurements collected by \citet{stickland:1984}, we obtained $P_{\rm orb}=22.01196\pm0.00077$~d, $HJD_0=2430707.97\pm0.48$, $K_1=48.73\pm0.91$~\kms, $K_2=89.3\pm3.9$~\kms, $\gamma=22.22\pm0.72$~\kms, $e=0.342\pm0.018$, and $\omega=104.2\pm3.9\degr$. The resulting orbital fit is shown in Fig.~\ref{fig:rv4}. Spectroscopic $P_{\rm orb}$ and $e$ agree with the information obtained from the TESS light curve.

\begin{figure}
\centering
\figps{0.81\hsize}{0}{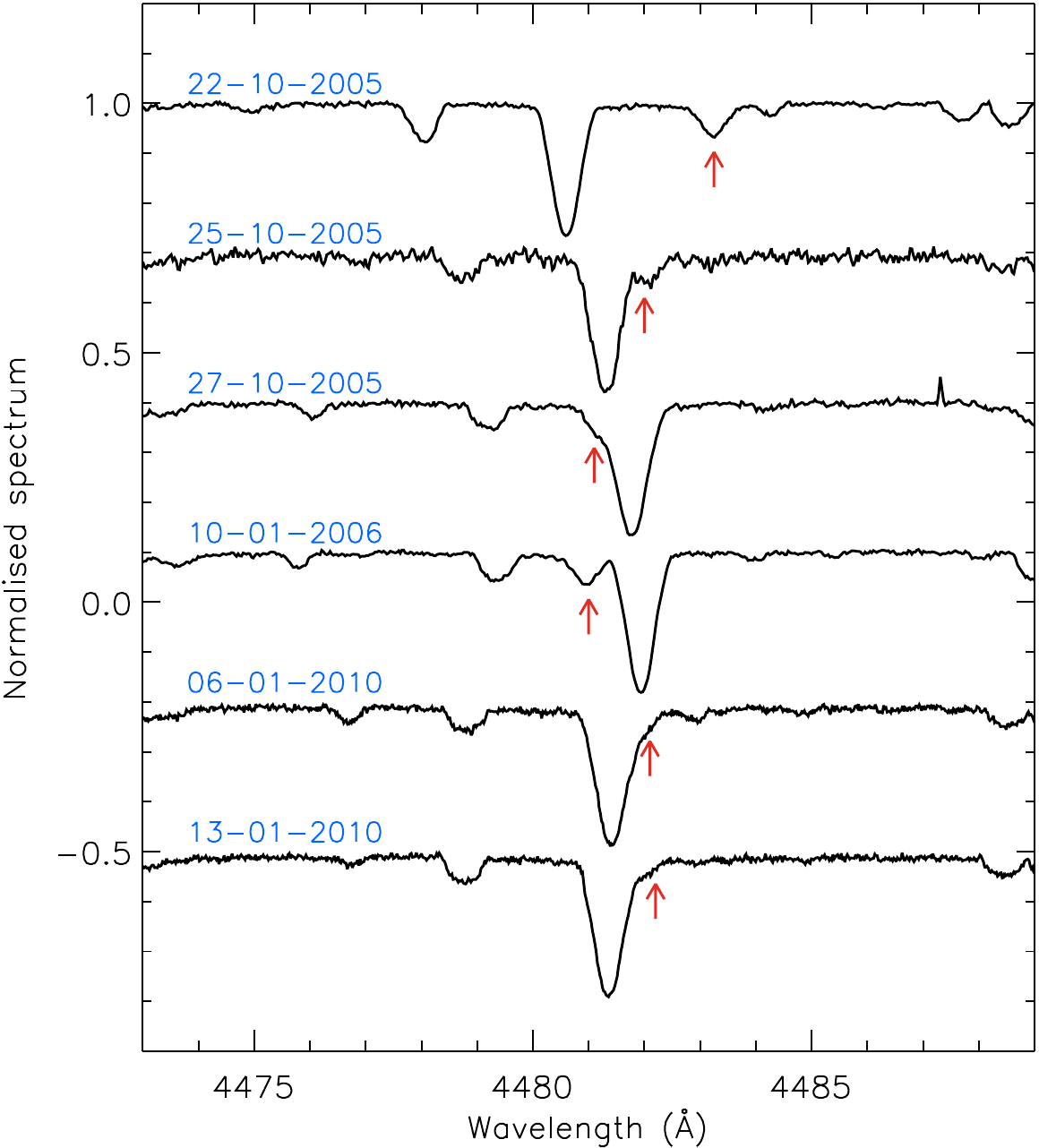}
\caption{Spectra of HD\,72208 in the vicinity of the Mg~{\sc ii} 4481~\AA\ line. Arrows indicate the contribution of the secondary. Spectra are offset vertically for display purpose. UT observing dates are indicated to the left.}
\label{fig:spec}
\end{figure}

The radial velocities provided by \citet{stickland:1984} were derived from low-quality photographic spectra, which allowed detection of secondary's lines in a small subset of observations. We examined modern high-resolution FEROS and HARPS observations of HD\,72208 available in the ESO archive\footnote{\url{http://archive.eso.org/eso/eso_archive_main.html}}. These spectra in the vicinity of the Mg~{\sc ii} 4481~\AA\ line are shown in Fig.~\ref{fig:spec}. The contribution of the secondary is clearly detected in all six spectra. 

Atmospheric parameters and abundances of the primary component of HD\,72208 were investigated by \citet{woolf:1999}. They reported $T_{\rm eff}=10900$~K, $\log g=3.87$, $v_{\rm e}\sin i=24.2$~\kms\ and measured abundances of Cr, Fe, and Hg. In another study \citet{makaganiuk:2011a} were unable to detect a longitudinal magnetic field stronger than about 100~G using two HARPSpol observations. \citet{scholler:2010} reported a close visual companion to HD\,72208. However, with a magnitude difference of more than 6~mag in K band, it contributes negligibly to the total light of the system and cannot be responsible for the eclipses in the 22~d binary.

\subsection{Rejected candidates}

The following targets showed eclipse variability in QLP light curves but were rejected by our pixel-by-pixel analysis: TYC\,1875-2341-1 (TIC 78971359), HD\,338483 (TIC 112504287), HD\,52310 (TIC 268459947), TYC\,3583-933-1 (TIC 366150938), TYC\,4066-400-1 (TIC 390897856), TYC\,4063-469-1 (TIC 392777979), HD\,71833 (TIC 412961700), TYC\,4047-2079-1 (TIC 49945382). For all these stars EB signals were offset from the positions of the HgMn objects, suggesting contamination from close neighbours. In addition, for TYC\,2859-368-1 (TIC 418082430), which is located in a crowded region, we were unable to find any EB signal on target or in its immediate vicinity that would resemble a very clear 2.867~d eclipse variation seen in the QLP light curve. Lack of radial velocity variation reported for this star by \citet{chojnowski:2020} confirms that it is not a short-period binary.

\section{Discussion}
\label{sec:disc}

We reported observation of eclipses in six binaries with HgMn components. This doubles the number of such chemically peculiar stars known in EBs. We confirmed the presence of eclipses in the bright SB2 star HD\,72208, which becomes the longest-period EB with an HgMn component. Including this object, only two other well-established (HD\,34364, TYC 455-791-1) and one candidate (HD\,161701) HgMn stars are double-line binaries with both components accessible for radial velocity measurements and spectroscopic analysis. 

Two systems studied here, HD\,36892 and 53004, exhibit heartbeat variations only recently discovered in HgMn binaries \citep{paunzen:2021a}. The latter star is also the first example of an HgMn star with multi-periodic pulsations excited by tidal perturbations in an eccentric short period binary \citep[e.g.][]{thompson:2012}. Furthermore, HD\,36892, 50984, and TYC 4047-570-1 show ellipsoidal variation due to distorted component shapes \citep{kochukhov:2021a}.

We demonstrated that existing radial velocity measurements of HD\,36892, 50984, 55776, and 72208 are compatible with the orbital periods derived from TESS photometry. For two of these stars, HD\,36892 and 55776, we corrected previously reported spurious spectroscopic orbital solutions. HD\,72208 and 53004 standout from the group of stars studied here owing to their brightness. We plan to carry out spectroscopic follow-up of these systems with the goal of performing complete characterisations of the components and measuring their surface abundances.

\section*{Acknowledgements}
O.K. acknowledges support by the Swedish Research Council, the Royal Swedish Academy of Sciences and the Swedish National Space Agency.
V.K. acknowledges support from the Natural Sciences and Engineering Research Council of Canada (NSERC) and from the Facult\'{e} des \'{E}tudes Sup\'{e}rieures et de la Recherch de l'Universit\'{e} de Moncton.
J.L.-B. acknowledges support from FAPESP (grant 2017/23731-1).
M.E.S. acknowledges support from the Annie Jump Cannon Fellowship, supported by the University of Delaware and endowed by the Mount Cuba Astronomical Observatory.
This paper includes data collected by the TESS mission, 
which is funded by NASA's Science Mission directorate.
This research made use of Lightkurve, a Python package for Kepler and TESS data analysis \citep{Lightkurve2018},
and of the SIMBAD database, operated at CDS, Strasbourg, France.

\section*{Data availability}
The TESS light curve data underlying this article can be obtained from the Mikulski Archive for Space Telescopes (MAST)\footnote{\url{https://mast.stsci.edu}}.


\bsp
\label{lastpage}
\end{document}